\documentclass[12pt]{article}
\usepackage{amsmath,amssymb,amsthm,amsxtra,overpic,bbm,bm,epsfig,subfigure}
\usepackage{hyperref}
\usepackage{mathrsfs}
\usepackage{graphicx}
\usepackage{color}
\usepackage{comment}
\usepackage{float}
\usepackage{cite}
\usepackage{tikz}

\usepackage{slashed,stmaryrd}

\def\thefootnote{\fnsymbol{footnote}}

\usepackage{quantikz}  
\def\gateO{\gate[style={fill=black},label style=white]{|{\bf 0}\rangle}}
\def\gateH{\gate[style={fill=cyan!80},label style=black]{\bf H}}
\def\gateX{\gate[style={fill=olive!50},label style=black]{\bf X}}
\def\gateU#1{\gate[style={fill=magenta!80},label style=black]{{\bf U}(#1)}}

\def\gateRy#1{\gate[style={fill=magenta!80},label style=black]{{\bf Ry}(#1)}}
\def\gateRz#1{\gate[style={fill=magenta!80},label style=black]{{\bf Rz}(#1)}}
\def\meas#1{\meter[fill=black,path picture={\draw[white] ([shift={(.1,.24)}]path picture bounding box.south west) to[bend left=50] ([shift={(-.1,.24)}]path picture bounding box.south east);\draw[-{Latex[scale=0.6]},white] ([shift={(0,.1)}]path picture bounding box.south) -- ([shift={(.3,-.1)}]path picture bounding box.north);}]{#1}}

\begin{document}

\vspace{0.2cm}

\begin{center}
{\Large\bf Quantum Simulations of the Non-Unitary Time Evolution and Applications to Neutral-Kaon Oscillations}
\end{center}

\vspace{0.2cm}

\begin{center}
{\bf Ying Chen}~$^{a,~b}$~\footnote{E-mail: cheny@ihep.ac.cn},
\quad
{\bf Yunheng Ma}~$^{a,~b}$~\footnote{E-mail: mayunheng@ihep.ac.cn (corresponding author)},
\quad
{\bf Shun Zhou}~$^{a,~b}$~\footnote{E-mail: zhoush@ihep.ac.cn (corresponding author)}
\\
\vspace{0.2cm}
{\small $^a$Institute of High Energy Physics, Chinese Academy of Sciences, Beijing 100049, China\\
$^b$School of Physical Sciences, University of Chinese Academy of Sciences, Beijing 100049, China}
\end{center}

\vspace{1.5cm}

\begin{abstract}
In light of recent exciting progress in building up quantum computing facilities based on both optical and cold-atom techniques, the algorithms for quantum simulations of particle-physics systems are in rapid progress. In this paper, we propose an efficient algorithm for simulating the non-unitary time evolution of neutral-kaon oscillations $K^0 \leftrightarrow \overline{K}^0$, with or without CP conservation, on the quantum computers provided by the IBM company. The essential strategy is to realize the time-evolution operator with basic quantum gates and an extra qubit corresponding to some external environment. The final results are well consistent with theoretical expectations, and the algorithm can also be applied to open systems beyond elementary particles.
\end{abstract}

\newpage

\def\thefootnote{\arabic{footnote}}
\setcounter{footnote}{0}

\section{Introduction}\label{sec:intro}

Since the concept of quantum computation was born in the early 1980's\cite{Benioff:1980, Manin:1980, Feynman:1981tf, Benioff:1982}, there has been tremendous progress in the formulation of quantum algorithms~\cite{Deutsch:1985, Deutsch:1989, Shor:1994, Grover:1996, Shor:1997, Zalka:1996st, Wiesner:1996xg, Bernstein:1997, Boghosian:1996qd, Abrams:1997gk, Preskill:1997ds, Jordan:2011ne} and the actual implementation of quantum computers~\cite{Chuang:1997, Cory, Chuang:2001, Blatt, Aspuru, Houck, Gross, Yang:2020yer}. As an interdisciplinary research area of computer science and physics~\cite{book}, quantum computation has recently attracted a lot attention from the communities of particle physics and nuclear physics and has realized various interesting applications~\cite{Jordan:2011ci, Jordan:2014tma, Preskill:2018fag, Arguelles:2019phs, Roggero:2019myu, Shaw:2020udc, Haase:2020kaj, Dasgupta:2020itb, Zhang:2020uqo, Ott:2020ycj, Atas:2021ext, Hall:2021rbv, Rahman:2021yse, Roggero:2021fyo, Zache:2021lrh, Davoudi:2021ney, Yeter-Aydeniz:2021olz}. Although the implementation of quantum computers to solve a realistic problem that cannot be easily treated with ordinary computers is still in the preliminary stage due to the limited number of qubits, we have seen rapidly increasing interest and enormous efforts in this area in the past few decades and thus expect groundbreaking scientific discoveries based on quantum computations in the near future~\cite{Pre18, Alexeev:2019enj}.

As one of the simplest and most important phenomena in particle physics, neutrino oscillations have been thoroughly studied in Ref.~\cite{Arguelles:2019phs} by using quantum processors. It has been demonstrated that the probabilities of two- and three-flavor neutrino oscillations, either with or without matter effects and extra sterile neutrino species, can be calculated by simulations on the classical computer and by the quantum hardware. In the present work, we are well motivated to examine another equally important phenomenon, namely, neutral-kaon oscillations $K^0 \leftrightarrow \overline{K}^0$~\cite{GellMann:1955jx}, and calculate the oscillation probabilities via quantum simulations. The primary motivation for such an examination is two-fold. First, the time evolution of a neutrino flavor eigenstate is actually governed by the Schr\"{o}dinger equation with a Hermitian Hamiltonian, implying that the time-evolution operator is unitary. In contrast, the effective Hamiltonian for the neutral-kaon system in the Weisskopf-Wigner approximation is given by~\cite{Weisskopf:1930au, Lee:1957qq}
\begin{eqnarray}
{\sf H} = {\sf M} - \frac{{\rm i}}{2} {\sf \Gamma} \; ,
\label{eq:hamiltonian}
\end{eqnarray}
where ${\sf H}$ denotes a general $2\times 2$ complex matrix, while ${\sf M}$ and ${\sf \Gamma}$ are $2\times 2$ Hermitian matrices. Since the effective Hamiltonian ${\sf H}$ is no longer Hermitian, the non-unitary time evolution of the quantum state of kaon cannot be realized in a straightforward way by the basic unitary gates in the context of quantum circuits. Second, the non-conservation of the charge-parity (CP) symmetry has been observed in the neutral-kaon system~\cite{Christenson:1964fg}, although the CP violation is rather small, the time evolution of the initial $K^0$ state and that of the $\overline{K}^0$ state are in  principle distinguishable. This provides us with a well-motivated example to investigate the non-unitary evolution with CP violation in a two-level quantum system at quantum computers.

In fact, the neutral-kaon oscillations serve as a simple open system,\footnote{Although we concentrate on the neutral-kaon system and restrict discussions to this particular case, other neutral-meson systems (e.g., $D^0$-$\overline{D}^0$ and $B^0$-$\overline{B}^0$~\cite{Branco:1999fs, Bigi:2000yz, Zyla:2020zbs}) can be discussed in a similar way.}  because both $K^0$ and $\overline{K}^0$ decay quickly into other light particles, which have not been taken into account in the effective Hamiltonian in the Weisskopf-Wigner approximation. For open systems, there exist interactions between the system in question and the environment~\cite{Breuer:2002pc, Riv12}, so the time evolution of the concerned system can hardly be described by a unitary operator. Looking for the solutions to open quantum systems is an advanced topic, where the Lindblad master equation~\cite{Lindblad:1975ef, Gorini:1975nb, Franke:1976, Weinberg:2015}and the Kraus representation~\cite{Kra83, Alicki:2007} are involved. The great advantage of quantum computing techniques may shed some light on the ultimate solutions to open systems. Along this line, the non-unitary quantum circuits have been previously studied in Refs.~\cite{Ter05, Sa:2021dns, Gup20}. As will be shown later, we propose a distinct and efficient algorithm for the quantum simulations of neutral-kaon oscillations.

The remaining part of the present paper is organized as follows. In Sec.~\ref{sec:osc}, in order to establish our notations, we briefly recall the description of the neutral-kaon system and recapitulate the results of oscillation probabilities. The realization of the non-unitary time evolution in the model of quantum circuits will be explained in Sec.~\ref{sec:quantum}, where we also use the quantum device from IBM~\cite{ibm} to demonstrate our strategy. The impact of CP violation on the algorithm is discussed. Finally, we summarize our main conclusions in Sec.~\ref{sec:sum}, and provide some necessary information about the IBM quantum computer and useful calculational details in the Appendix.

\section{Neutral-Kaon Oscillations}\label{sec:osc}

As the first example for the particle-antiparticle mixing phenomenon, the neutral-meson system has been well studied for a long time both theoretically~\cite{Bigi:2000yz} and experimentally~\cite{Zyla:2020zbs}, so we just recapitulate the essential ingredients relevant for our subsequent discussions. In the flavor basis $|K^0\rangle \equiv (1, 0)^{\rm T}$ and $|\overline{K}^0\rangle \equiv (0, 1)^{\rm T}$, the time evolution of the quantum state $|K(t)\rangle \equiv c(t) |K^0\rangle + \overline{c}(t) |\overline{K}^0\rangle$, where the functions $c(t)$ and $\overline{c}(t)$ stand respectively for the probability amplitudes in the $|K^0\rangle$ and $|\overline{K}^0\rangle$ state, is described by the Schr\"{o}dinger equation
\begin{eqnarray}
{\rm i} \frac{\rm d}{{\rm d}t} \left( \begin{matrix} c(t) \cr \overline{c}(t)\end{matrix}\right) = \left[ \left( \begin{matrix} M^{}_{11} & M^{}_{12} \cr M^*_{12} & M^{}_{22} \end{matrix} \right) - \frac{\rm i} {2} \left( \begin{matrix} \Gamma^{}_{11} & \Gamma^{}_{12} \cr \Gamma^*_{12} & \Gamma^{}_{22} \end{matrix} \right) \right]\left( \begin{matrix} c(t) \cr \overline{c}(t)\end{matrix}\right) \; ,
\label{eq:schrodinger}
\end{eqnarray}
with $M^{}_{ij}$ and $\Gamma^{}_{ij}$ (for $i, j = 1, 2$) being the elements of the Hermitian matrices ${\sf M}$ and ${\sf \Gamma}$ in the effective Hamiltonian introduced in Eq.~(\ref{eq:hamiltonian}). The joint CP and time-reversal (CPT) symmetry requires $M^{}_{11} = M^{}_{22} \equiv M$ and $\Gamma^{}_{11} = \Gamma^{}_{22} \equiv \Gamma$, which will always be assumed in the present work. However, since CP violation has been experimentally observed in the neutral-kaon system~\cite{Christenson:1964fg}, the off-diagonal elements $M^{}_{12}$ and $\Gamma^{}_{12}$ must be complex.

As in the usual way, given the initial state $|K(0)\rangle = |K^0\rangle$ at $t = 0$ or equivalently $c(0)=1$ and $\overline{c}(0) = 0$, the transition $K^0 \to \overline{K}^0$ and survival $K^0 \to K^0$ probabilities can be calculated by solving the Schr\"{o}dinger equation in Eq.~(\ref{eq:schrodinger}). This is usually achieved by first diagonalizing the effective Hamiltonian ${\sf H}$ via a similarity transformation ${\sf A}\,{\sf H}\,{\sf A}^{-1} = {\rm Diag}\{E^{}_{\rm S}, E^{}_{\rm L}\}$ with the complex eigenvalues $E^{}_{\rm S} = M^{}_{\rm S} - {\rm i}\,\Gamma^{}_{\rm S}/2$ and $E^{}_{\rm L} = M^{}_{\rm L} - {\rm i}\,\Gamma^{}_{\rm L}/2$, where $M^{}_{\rm S, L}$ and $\Gamma^{}_{\rm S, L}$ denote the masses and decay widths of the corresponding energy eigenstates $|K^{}_{\rm S}\rangle$ and $|K^{}_{\rm L}\rangle$, respectively. Since the Hamiltonian ${\sf H}$ is non-Hermitian, the transformation matrix ${\sf A}$ is not unitary any more. Without repeating the standard treatment, we just summarize the main results below.
\begin{itemize}
\item The energy eigenstates $|K^{}_{\rm S}\rangle$ and $|K^{}_{\rm L}\rangle$ can be expressed in terms of $|K^0\rangle$ and $|\overline{K}^0\rangle$ with the help of the transformation matrix ${\sf A}$, namely,
\begin{eqnarray}
|K^{}_{\rm S}\rangle &=& p |K^0\rangle + q |\overline{K}^0\rangle \; , \nonumber \\
|K^{}_{\rm L}\rangle &=& p |K^0\rangle - q |\overline{K}^0\rangle  \; ,
\label{eq:KSKL}
\end{eqnarray}
where the complex parameters $p$ and $q$ are related to the matrix elements $M^{}_{12}$ and $\Gamma^{}_{12}$ as $p^2 \equiv M^{}_{12} - {\rm i}\,\Gamma^{}_{12}/2$ and $q^2 \equiv M^*_{12} - {\rm i}\,\Gamma^*_{12}/2$, and the normalization condition $|p|^2 + |q|^2 = 1$ is adopted~\cite{Zyla:2020zbs}. The corresponding masses and decay widths are given by
\begin{eqnarray}
M^{}_{\rm S,L} &=& M \mp {\rm Re}\left[\sqrt{|M^{}_{12}|^2 - |\Gamma^{}_{12}|^2/4 - {\rm i}\,{\rm Re}\left(M^{}_{12} \Gamma^*_{12}\right)}\right] \; , \nonumber \\
\Gamma^{}_{\rm S,L} &=& \Gamma \pm 2\,{\rm Im}\left[\sqrt{|M^{}_{12}|^2 - |\Gamma^{}_{12}|^2/4 - {\rm i}\,{\rm Re}\left(M^{}_{12} \Gamma^*_{12}\right)}\right] \; ,
\label{eq:masswidth}
\end{eqnarray}
where the upper and lower signs refer to the cases of $|K^{}_{\rm S}\rangle$ and $|K^{}_{\rm L}\rangle$, respectively. Experimentally the long-lived neutral meson $K^{}_{\rm L}$ is found to be heavier than the short-lived one $K^{}_{\rm S}$, implying $\Delta m = M^{}_{\rm L} - M^{}_{\rm S} > 0$ and $\Delta \Gamma \equiv \Gamma^{}_{\rm L} - \Gamma^{}_{\rm S} < 0$. In addition, from Eq.~(\ref{eq:masswidth}), we can obtain the average mass $M = (M^{}_{\rm S} + M^{}_{\rm L})/2$ and decay width $\Gamma = (\Gamma^{}_{\rm L} + \Gamma^{}_{\rm S})/2$. It is easy to verify that $\Delta m \cdot \Delta \Gamma = 4\,{\rm Re}(M^{}_{12} \Gamma^*_{12})$ and $(\Delta m)^2 - (\Delta \Gamma)^2/4 = 4 (|M^{}_{12}|^2 - |\Gamma^{}_{12}|^2/4)$ are exactly valid.

According to the latest experimental measurements of the masses and decay widths~\cite{Zyla:2020zbs}, one can immediately get $M = 497.6~{\rm MeV}$, $\Gamma =  3.691\times 10^{-12}~{\rm MeV}$, $\Delta m = 3.494\times 10^{-12}~{\rm MeV}$, and $\Delta \Gamma = -7.356\times 10^{-12}~{\rm MeV}$. In the presence of CP violation, i.e., ${\rm Im}(M^{}_{12}) \neq 0$ and ${\rm Im}(\Gamma^{}_{12}) \neq 0$, there exist five real parameters relevant for the neutral-kaon system, namely, $M$, $\Gamma$, $M^{}_{12} \equiv |M^{}_{12}| e^{{\rm i}\phi^{}_{\rm M}}$ and $\Gamma^{}_{12} \equiv |\Gamma^{}_{12}|e^{{\rm i}\phi^{}_{\rm \Gamma}}$, where only the phase difference $\phi^{}_{\rm M} - \phi^{}_{\rm \Gamma}$ is physical due to the freedom of redefining the phases of the states $|K^0\rangle$ and $|\overline{K}^0\rangle$~\cite{Branco:1999fs}. If the direct CP violation is ignored, such a phase difference can be unambiguously extracted from the CP asymmetry observed in the semi-leptonic decays of $K^{}_{\rm L}$, i.e.,
\begin{eqnarray}
A^{}_{\rm L} \equiv \frac{\Gamma(K^{}_{\rm L} \to \pi^- \ell^+ \nu^{}_\ell) - \Gamma(K^{}_{\rm L} \to \pi^+ \ell^- \overline{\nu}^{}_\ell)}{\Gamma(K^{}_{\rm L} \to \pi^- \ell^+ \nu^{}_\ell) + \Gamma(K^{}_{\rm L} \to \pi^+ \ell^- \overline{\nu}^{}_\ell)} = \frac{1 - |q/p|^2}{1 + |q/p|^2} \; ,
\label{eq:AL}
\end{eqnarray}
with $\ell = e$ or $\mu$. The weighted average of $A^{}_{\rm L}(e)$ and $A^{}_{\rm L}(\mu)$ has been reported by Particle Data Group and the best-fit value is quoted as $A^{}_{\rm L} = 3.32\times 10^{-3}$~\cite{Zyla:2020zbs}. Given this tiny asymmetry, one can observe $|q/p|^2 = (1 - A^{}_{\rm L})/(1 + A^{}_{\rm L}) \approx 1$ and thus obtain $|M^{}_{12}| \approx \Delta m/2 = 1.747\times 10^{-12}~{\rm MeV}$, $|\Gamma^{}_{12}| \approx -\Delta \Gamma/2 = 3.678\times 10^{-12}~{\rm MeV}$ and $\cos(\phi^{}_{\rm M} - \phi^{}_\Gamma) \approx -1$. From the definitions of $q^2$ and $p^2$, together with Eq.~(\ref{eq:AL}), we can determine the phase difference via the relation
\begin{eqnarray}
\sin(\phi^{}_{\rm M} - \phi^{}_{\rm \Gamma}) = - \frac{2A^{}_{\rm L}}{1 + A^2_{\rm L}} \cdot \frac{|M^{}_{12}|^2 + |\Gamma^{}_{12}|^2/4}{|M^{}_{12}| |\Gamma^{}_{12}|} \approx -6.649\times 10^{-3} \; ,
\label{eq:phase}
\end{eqnarray}
indicating $\phi^{}_{\rm M} - \phi^{}_\Gamma \approx 180.4^\circ$. Thus far all the five real parameters in the neutral-kaon system have been fixed by experimental observations.

\item Since the time evolution of the energy eigenstates $|K^{}_{\rm S}\rangle$ and $|K^{}_{\rm L}\rangle$ will be simply described by the phase factors $\exp(-{\rm i} E^{}_{\rm S}t)$ and $\exp(-{\rm i} E^{}_{\rm L}t)$, respectively, the physical neutral-kaon states at the elapsed time $t$ can be obtained by transforming back to the $|K^0\rangle$-$|\overline{K}^0\rangle$ basis~\cite{Zyla:2020zbs}, i.e.,
\begin{eqnarray}
|K^0_{\rm phys}(t)\rangle &=& g^{}_+(t) |K^0\rangle - \frac{q}{p} g^{}_-(t) |\overline{K}^0\rangle \; , \nonumber \\
|\overline{K}^0_{\rm phys}(t)\rangle &=& g^{}_+(t) |\overline{K}^0\rangle - \frac{p}{q} g^{}_-(t) |K^0\rangle\; ,
\label{eq:Kt}
\end{eqnarray}
where $g^{}_\pm(t) \equiv \left[\exp(-{\rm i}E^{}_{\rm L}t) \pm \exp(-{\rm i} E^{}_{\rm S}t) \right]/2$. If the initial state is $|K^0\rangle$, then the survival and transition probabilities can be derived straightforwardly as
\begin{eqnarray}
P(K^0 \to K^0) &=& \left|g^{}_+(t)\right|^2 = \frac{1}{4}\left[e^{-\Gamma^{}_{\rm S}t} + e^{-\Gamma^{}_{\rm L}t} + 2 e^{-\Gamma t} \cos(\Delta m t)\right] \;, \nonumber \\
P(K^0 \to \overline{K}^0) &=& \left|\frac{q}{p}\right|^2 \left|g^{}_-(t)\right|^2 = \frac{1}{4} \left|\frac{q}{p}\right|^2  \left[e^{-\Gamma^{}_{\rm S}t} + e^{-\Gamma^{}_{\rm L}t} - 2 e^{-\Gamma t} \cos(\Delta m t)\right]\; ,
\label{eq:prob}
\end{eqnarray}
where the dissipative effects in the system due to particle decays are clearly represented by the overall exponential functions, while the oscillatory behavior is dictated by the cosine function. Similarly, for the initial state $|\overline{K}^0\rangle$, the survival probability remains the same $P(\overline{K}^0 \to \overline{K}^0) = |g^{}_+(t)|^2 = P(K^0 \to K^0)$, as it should be because of the CPT symmetry, while the transition probability turns out to be $P(\overline{K}^0 \to K^0) = |p/q|^2 \cdot \left|g^{}_-(t)\right|^2  = |p/q|^4 \cdot P(K^0 \to \overline{K}^0)$. Therefore, the CP asymmetry in the neutral-kaon oscillations is
\begin{eqnarray}
A^{}_{\rm CP} &\equiv& \frac{P(\overline{K}^0 \to K^0) - P(K^0 \to \overline{K}^0) }{P(\overline{K}^0 \to K^0) + P(K^0 \to \overline{K}^0)} = \frac{1 - |q|^4/|p|^4}{1 + |q|^4/|p|^4} \approx 2 A^{}_{\rm L} \; ,
\label{eq:ACP}
\end{eqnarray}
which signifies the indirect CP violation from the $K^0$-$\overline{K}^0$ mixing. As one can see, this CP asymmetry $A^{}_{\rm CP} \approx 6.64\times 10^{-3}$ is highly suppressed such that it is difficult to discriminate $P(\overline{K}^0 \to K^0)$ from $P(K^0 \to \overline{K}^0)$.
\end{itemize}

It is worth mentioning that the neutral kaons will decay quickly into two or three pions in reality, which cannot be accounted for by the effective Hamiltonian that has been restricted into the two-dimensional Hilbert space spanned only by $|K^0\rangle$ and $|\overline{K}^0\rangle$. The inclusion of particle decays in quantum computations will be interesting on its own and deserve further exploration~\cite{Ciavarella:2020vqm}.

\section{Quantum Simulations}\label{sec:quantum}

Roughly speaking, the quantum computer is a computing device that makes use of quantum effects of some optical or atomic system. Physical operations on a quantum state of such a system play essentially the same role of basic logical gates acting on qubits. Therefore, the main task of quantum algorithms is to decompose an operator into a series of basic logical gates.

In our case, we have to express the time-evolution operator for the $K^0$-$\overline{K}^0$ system in terms of some basic logical gates. Such an expression can be translated into a quantum circuit that always ends with the measurements in the ${\bf Z}$ basis~\cite{book}. The measurement on this circuit corresponds to a real observation on the implemented physical system. With a large number of measurements, one can construct the transition probabilities for a given initial state of our interest. More explicitly, we perform quantum simulations to infer the survival and transition probabilities for the neutral-kaon system starting with the $|K^0\rangle$ state.

\subsection{CP Conservation}\label{subsec:CPC}

If CP conservation is assumed, the matrix elements $M^{}_{12}$ and $\Gamma^{}_{12}$ are real, namely, ${\rm Im}(M^{}_{12}) = {\rm Im}(\Gamma^{}_{12}) = 0$. In this case, the effective Hamiltonian in Eq.~(\ref{eq:hamiltonian}) can be recast into the following form
\begin{eqnarray}
{\sf H} = \left(M - \frac{\rm i}{2} \Gamma\right) {\bf I} + \left(M^{}_{12} - \frac{\rm i}{2} \Gamma^{}_{12}\right) {\bm \sigma}^{}_x \; ,
\label{eq:sigmaform}
\end{eqnarray}
where ${\bf I}$ denotes the two-dimensional identity matrix and ${\bm \sigma}^{}_x$ is the first Pauli matrix. Since ${\bf I}$ and ${\bm \sigma}^{}_x$ commute with each other, the time-evolution operator can be decomposed into
\begin{eqnarray}
{\sf E}(t) \equiv \exp(-{\rm i} {\sf H} t) = \prod^4_{i = 1} {\sf E}^{}_i(t) \; ,
\label{eq:product}
\end{eqnarray}
where the evolution matrices ${\sf E}^{}_i(t)$ can be easily found as
\begin{eqnarray}
{\sf E}^{}_1(t) = e^{-{\rm i} M t} \, {\bf I} \; , \quad {\sf E}^{}_2(t) = e^{- \Gamma t/2}\, {\bf I} \; ,
\label{eq:E1E2}
\end{eqnarray}
and
\begin{eqnarray}
{\sf E}^{}_3(t) = \left(\begin{matrix}\cos{\left(M^{}_{12} t \right)} & - {\rm i} \sin{\left(M^{}_{12} t \right)}\\- {\rm i} \sin{\left(M^{}_{12} t \right)} & \cos{\left(M^{}_{12} t \right)}\end{matrix}\right) \; , \quad {\sf E}^{}_4(t) = \left(\begin{matrix}\cosh{\left(\Gamma^{}_{12} t/2 \right)} & - \sinh{\left(\Gamma^{}_{12} t/2 \right)} \\ - \sinh{\left(\Gamma^{}_{12} t/2 \right)} & \cosh{\left(\Gamma^{}_{12} t/2\right)}\end{matrix}\right) \; .
\label{eq:E3E4}
\end{eqnarray}
The symmetric matrices ${\sf E}^{}_3(t)$ and ${\sf E}^{}_4(t)$ can be diagonalized via the similarity transformations ${\sf A}\, {\sf E}^{}_{3,4}(t)\, {\sf A}^{-1} = \widehat{\sf E}^{}_{3,4}(t)$,  where the orthogonal matrix $\sf A$ is
\begin{align}
{\sf A} = \frac{1}{\sqrt 2}\left(\begin{matrix}1 &  1 \\ 1 & -1 \end{matrix}\right)\; ,
\label{eq:Amatrix}
\end{align}
and the diagonal matrices $\widehat{\sf E}_{3,4}$ are
\begin{eqnarray}
\widehat{\sf E}^{}_3(t) = \left(\begin{matrix}e^{-{\rm i}M^{}_{12}t} & 0 \\ 0 & e^{+{\rm i}M^{}_{12}t}\end{matrix}\right) \; , \quad \widehat{\sf E}^{}_4(t) = \left(\begin{matrix}e^{- \Gamma^{}_{12} t/2} & 0 \\ 0 & e^{+\Gamma^{}_{12} t/2}\end{matrix}\right) \; .
\label{eq:E3E4hat}
\end{eqnarray}
Noticing ${\sf A}^{-1} = {\sf A}^{\rm T} = {\sf A}$, the overall time-evolution operator ${\sf E(t)}$ is finally expressed as
\begin{eqnarray}
{\sf E}(t) = {\sf E}^{}_1(t) \cdot {\sf E}^{}_2(t) \cdot {\sf A} \cdot \widehat{\sf E}^{}_3(t) \cdot \widehat{\sf E}^{}_4(t) \cdot {\sf A} \; ,
\label{eq:finalE}
\end{eqnarray}
which has to be realized by using quantum circuits. Some helpful comments on Eq.~(\ref{eq:finalE}) are in order. First, one can recognize that ${\sf E}^{}_1(t)$ introduces a universal phase to both neutral-kaon states, which will be irrelevant for the oscillation probabilities. In contrast, the evolution matrix ${\sf E}^{}_2(t)$ leads to a global decay of the neutral-kaon system. Second, the evolution matrices $\widehat{\sf E}^{}_3(t)$ and $\widehat{\sf E}^{}_4(t)$ are lying between ${\sf A}^{-1}$ and ${\sf A}$, which implies a transformation from the flavor $K^0$-$\overline{K}^0$ basis to the energy $K^{}_{\rm S}$-$K^{}_{\rm L}$ basis. In the energy basis, $\widehat{\sf E}^{}_3(t)$ induces a relative phase between two energy eigenstates, implying an oscillatory behavior in the flavor basis. On the other hand, $\widehat{\sf E}^{}_4(t)$ shows that two energy eigenstates interact with the environment via the couplings of different strengths. As both ${\sf E}^{}_2(t)$ and $\widehat{\sf E}^{}_4(t)$ are obviously non-unitary for $t\neq 0$, the $K^0$-$\overline{K}^0$ system is an open quantum system that interacts with the environment of additional degrees of freedom. In fact, $\widehat{\sf E}^{}_4(t)$ drives the whole system into the preferred energy state $K^{}_{\rm L}$.

Now it is clear that the quantum simulation of the time evolution requires a quantum circuit to encode ${\sf E}(t)$. To this end, we first represent the flavor states $|K^0\rangle$ and $|\overline{K}^0\rangle$ by one qubit. Denoting the qubit state as $|q^{}_{\rm s}\rangle$, one can identify the two qubit states as $|0^{}_{\rm s}\rangle = |K^0\rangle = (1, 0)^{\rm T}$ and $|1^{}_{\rm s} \rangle = |\overline{K}^0\rangle = (0, 1)^{\rm T}$. In this way, a general quantum state of the $K^0$-$\overline{K}^0$ system can be represented by the qubit state $|q^{}_{\rm s}\rangle$ on a quantum computer. Then, we have to find quantum gates to encode ${\sf E}^{}_1(t)$, ${\sf E}^{}_2(t)$, $\widehat{\sf E}^{}_3(t)$, $\widehat{\sf E}_4(t)$ and ${\sf A}$ that appear in the evolution operator ${\sf E}(t)$. Our strategy to do so is outlined as follows.
\begin{table}[!t]
  \centering
  \begin{tabular}{ccc}
  \hline
    Gate & Symbol & Matrix \\ \hline
    {\bf H}
    &
    \begin{minipage}[b]{0.2\columnwidth}
		\centering
		\raisebox{-.4\height}{\includegraphics[width=\linewidth]{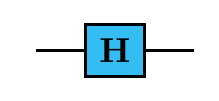}}
	\end{minipage}
    & $\displaystyle \frac{1}{\sqrt 2}
    \left(\begin{matrix}
    1 & 1 \cr
    1 &-1
    \end{matrix}\right)$ \vspace{0.2cm} \\
    {\bf X}
    &
    \begin{minipage}[b]{0.2\columnwidth}
		\centering
		\raisebox{-.4\height}{\includegraphics[width=\linewidth]{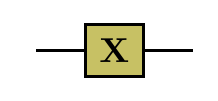}}
	\end{minipage}
    & $
    \left(\begin{matrix}
     0 & 1 \cr
    1 & 0
    \end{matrix}\right)$ \vspace{0.2cm} \\
    {\bf Rz}
    &
    \begin{minipage}[b]{0.2\columnwidth}
		\centering
		\raisebox{-.4\height}{\includegraphics[width=\linewidth]{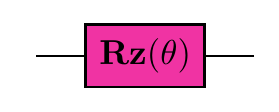}}
	\end{minipage}
    & $
    \left(\begin{matrix}
    e^{-{\rm i}\theta/2} & 0 \cr
    0 & e^{+{\rm i}\theta/2}
    \end{matrix}\right)$ \vspace{0.2cm} \\
    {\bf Cry}
    &
    \begin{minipage}[b]{0.2\columnwidth}
		\centering
		\raisebox{-.4\height}{\includegraphics[width=\linewidth]{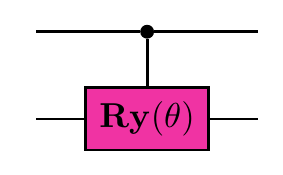}}
	\end{minipage}
    & $
    \left(\begin{matrix}
    1 & 0 & 0 & 0 \cr
    0 & 1 & 0 & 0 \cr
    0 & 0 & \cos \displaystyle \frac{\theta}{2} & -\sin \displaystyle \frac{\theta}{2} \cr
    0 & 0 & \sin \displaystyle \frac{\theta}{2} & \cos \displaystyle \frac{\theta}{2}
    \end{matrix}\right)$ \vspace{0.2cm} \\
    {\bf U}
    &
    \begin{minipage}[b]{0.25\columnwidth}
		\centering
		\raisebox{-.4\height}{\includegraphics[width=\linewidth]{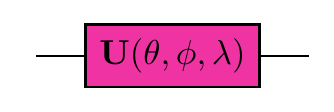}}
	\end{minipage}
    & $
    \left(\begin{matrix}
    \cos \displaystyle \frac{\theta}{2} & -e^{{\rm i}\lambda} \sin \displaystyle \frac{\theta}{2} \cr
    e^{{\rm i}\phi} \sin \displaystyle \frac{\theta}{2}& e^{{\rm i}\phi + {\rm i}\lambda} \cos \displaystyle \frac{\theta}{2}
    \end{matrix}\right)$
    \vspace{0.2cm} \\ \hline \hline
  \end{tabular}
  \vspace{0.2cm}
  \caption{Summary of the quantum gates used in this work, where we follow the notations from Ref.~\cite{book} and the symbols have been drawn by using the Quantikz package~\cite{Kay20}. For two-qubit gates, the computing bases are in order of $|00\rangle$, $|01\rangle$, $|10\rangle$, and $|11\rangle$.}
  \label{tab:gates}
\end{table}

First of all, as has been mentioned before, ${\sf E}^{}_1(t)$ does not affect the oscillation probabilities, so we just ignore it in our calculations. While $\widehat{\sf E}^{}_3(t)$ takes exactly the same matrix form of the quantum gate $\bf Rz(\theta)$ with $\theta = 2 M^{}_{12} t$, ${\sf A}$ is just the standard Hadamard gate ${\bf H}$. The quantum gates and their matrix forms have been summarized in Table~\ref{tab:gates}.

Second, since ${\sf E}^{}_2(t)$ is proportional to the identity matrix ${\bf I}$, it commutes with the matrix ${\sf A}$. Therefore, one can combine the non-unitary matrices ${\sf E}^{}_2(t)$ and $\widehat{\sf E}^{}_4(t)$ into
\begin{eqnarray}
{\sf N}^{}_0(t) = {\sf E}^{}_2(t) \cdot \widehat{\sf E}^{}_4(t) =
\left(\begin{matrix} e^{- (\Gamma + \Gamma^{}_{12})t/2} & 0 \\ 0 & e^{- (\Gamma - \Gamma^{}_{12})t/2} \end{matrix} \right) \equiv
\left(\begin{matrix} a(t) & 0 \\ 0 & b(t) \end{matrix}\right) \; ,
\label{eq:N0}
\end{eqnarray}
where $a(t) \equiv \exp \left[- (\Gamma + \Gamma^{}_{12})t/2 \right]$ and $b(t) = \exp \left[ - (\Gamma - \Gamma^{}_{12})t/2 \right]$ have been defined.  This operator is non-negative and non-unitary for $t>0$, so there is obviously no single qubit gate available to encode it.

In the Kraus representation theory~\cite{Kra83}, a non-unitary quantum operation on the principal system of interest can be viewed as a part of a unitary operation $\sf G$ on a larger closed system composed of the principal system and a complementary environment. In our case, the complementary environment can be realized by involving an additional qubit, whose state vector is denoted by $|q^{}_{\rm b}\rangle$ with the subscript referring to a thermal bath. Thus our main goal is to construct a unitary operator ${\sf G}$ acting on this two-qubit system. According to the Kraus theorem, there is no special requirement for the initial state of the complementary environment. Without loss of generality, we assume that the environment starts initially with the state $|0^{}_{\rm b}\rangle$. After the operation ${\sf G}$ on the joint system of $|q^{}_{\rm s}\rangle$ and $|q^{}_{\rm b}\rangle$, where the subscripts ``s" and ``b" correspond respectively to the left and right digits in the two-qubit computing basis, we apply the projection operator $|0^{}_{\rm b}\rangle \langle 0^{}_{\rm b}|$ to achieve the required operation, namely, ${\sf N}^{}_0(t) = \langle 0^{}_{\rm b}| {\sf G} |0^{}_{\rm b}\rangle$. There are many unitary matrices meeting this requirement, among which one convenient choice is
\begin{align}
{\sf G} = \left(\begin{matrix}
a&-\sqrt{1-a^2} & 0 & 0 \cr
\sqrt{1-a^2} & a & 0 & 0 \cr
0 & 0 & b & -\sqrt{1-b^2}\cr
0 & 0 & \sqrt{1-b^2}&b
\end{matrix}\right) \; ,
\label{eq:G}
\end{align}
where the argument $t$ of the functions $a(t)$, $b(t)$ and ${\sf G}(t)$ has been suppressed. Such a choice also implies another quantum operation when the projection by $|1^{}_{\rm b}\rangle \langle 1^{}_{\rm b}|$ is applied, namely,
\begin{align}
{\sf N}^{}_1(t) \equiv \langle 1^{}_{\rm b}| {\sf G} |0^{}_{\rm b}\rangle = \left(\begin{matrix}
\sqrt{1-a^2} & 0 \\ 0 & \sqrt{1-b^2}
\end{matrix}\right) \; ,
\label{eq:N1}
\end{align}
which actually corresponds to the evolution of the environment. Let's look closely into what ${\sf G}$ acutally does. If the system represented by $|q^{}_{\rm s}\rangle$ appears in an arbitrary state $|\psi^{}_{\rm i}\rangle = x|0^{}_{\rm s}\rangle + y|1^{}_{\rm s}\rangle$ and the environment $|q^{}_{\rm b}\rangle$ is initialized in $|0^{}_{\rm b}\rangle$, then the circuit state should be
\begin{align}
|\psi\rangle \equiv |\psi^{}_{\rm i}\rangle \otimes |0^{}_{\rm b}\rangle = x|0^{}_{\rm s}\rangle \otimes |0^{}_{\rm b}\rangle + y|1^{}_{\rm s}\rangle \otimes |0^{}_{\rm b}\rangle \; .
\label{eq:psi}
\end{align}
After the operation of ${\sf G}$, this state will be turned into
\begin{align}
|\psi^\prime\rangle = {\sf G}|\psi\rangle = a x |0^{}_{\rm s}\rangle \otimes |0^{}_{\rm b}\rangle + x \sqrt{1-a^2} |0^{}_{\rm s}\rangle \otimes |1^{}_{\rm b}\rangle + b y |1^{}_{\rm s} \rangle \otimes |0^{}_{\rm b}\rangle + y \sqrt{1-b^2} |1^{}_{\rm s}\rangle \otimes |1^{}_{\rm b}\rangle \; ,
\label{eq:psiprime}
\end{align}
from which one can recognize that the coefficient in front of $|0^{}_{\rm s} \rangle \otimes |0^{}_{\rm b}\rangle$ gives the required probability amplitude for the principal system to be in the $|K^0\rangle$ state, while that in front of $|1^{}_{\rm s}\rangle \otimes |0^{}_{\rm b}\rangle$ for the $|\overline{K}^0\rangle$ state. The leakage of the probability into the environment can also be calculated as
\begin{align}
1 - a^2|x|^2 - b^2|y|^2 = |x\sqrt{1-a^2}|^2 + |y\sqrt{1-b^2}|^2 \; .
\label{eq:leakage}
\end{align}
Therefore, we have demonstrated that ${\sf G}$ indeed gives rise to ${\sf N}^{}_0(t)$, when the environment state $|q^{}_{\rm b}\rangle$ ends in $|0^{}_{\rm b}\rangle$. If $|q^{}_{\rm b}\rangle$ ends in $|1^{}_{\rm b}\rangle$, we obtain the evolution of the environment. Now that ${\sf G}$ is unitary, it can be encoded by a series of quantum gates,
\begin{align}
{\sf G} \longrightarrow ({\bf X}\otimes {\bf I})\cdot {\bf Cry}(2\arccos a)\cdot ({\bf X}\otimes{\bf I})\cdot {\bf Cry}(2\arccos b) \; ,
\label{eq:Gcode}
\end{align}
where the matrix forms of the quantum gates ${\bf X}$ and ${\bf Cry}$ can be found in Table~\ref{tab:gates}.
\begin{figure}[!t]
	\begin{center}
    	\begin{tikzpicture}
        	\node[scale=0.95] {
            	\begin{quantikz}
                	\lstick{$q^{}_{\rm s}$} &\ket{0} &\gateH &\ctrl{1}\gategroup[wires=2,steps=4,style={dashed,rounded corners,fill=blue!20,inner xsep=2pt},background]{${\sf G}$} &\gateX &\ctrl{1}             &\gateX &\gateRz{2M_{12}t} &\gateH &\meas{} \\
                	\lstick{$q^{}_{\rm b}$} &        &\gateO    &\gateRy{2\arccos b}
   &\qw    &\gateRy{2\arccos a}  &\qw    &\qw               &\qw    &\meas{}
    	        \end{quantikz}
        	};
	    \end{tikzpicture}
	    \caption{Quantum circuit for the time evolution of the $K^0$-$\overline{K}^0$ system in the CP-conserving case, where the circuit is initialized in $|K^0\rangle$ state.}
	    \label{fig:cpc_circuit}
	\end{center}
\end{figure}
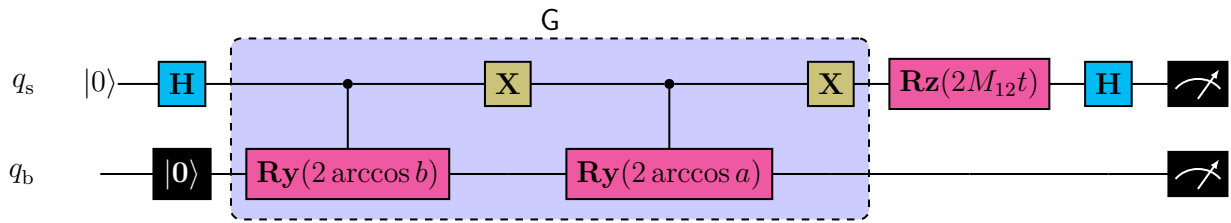

\begin{figure}[!t]
	\begin{center}
	\includegraphics[scale=0.65]{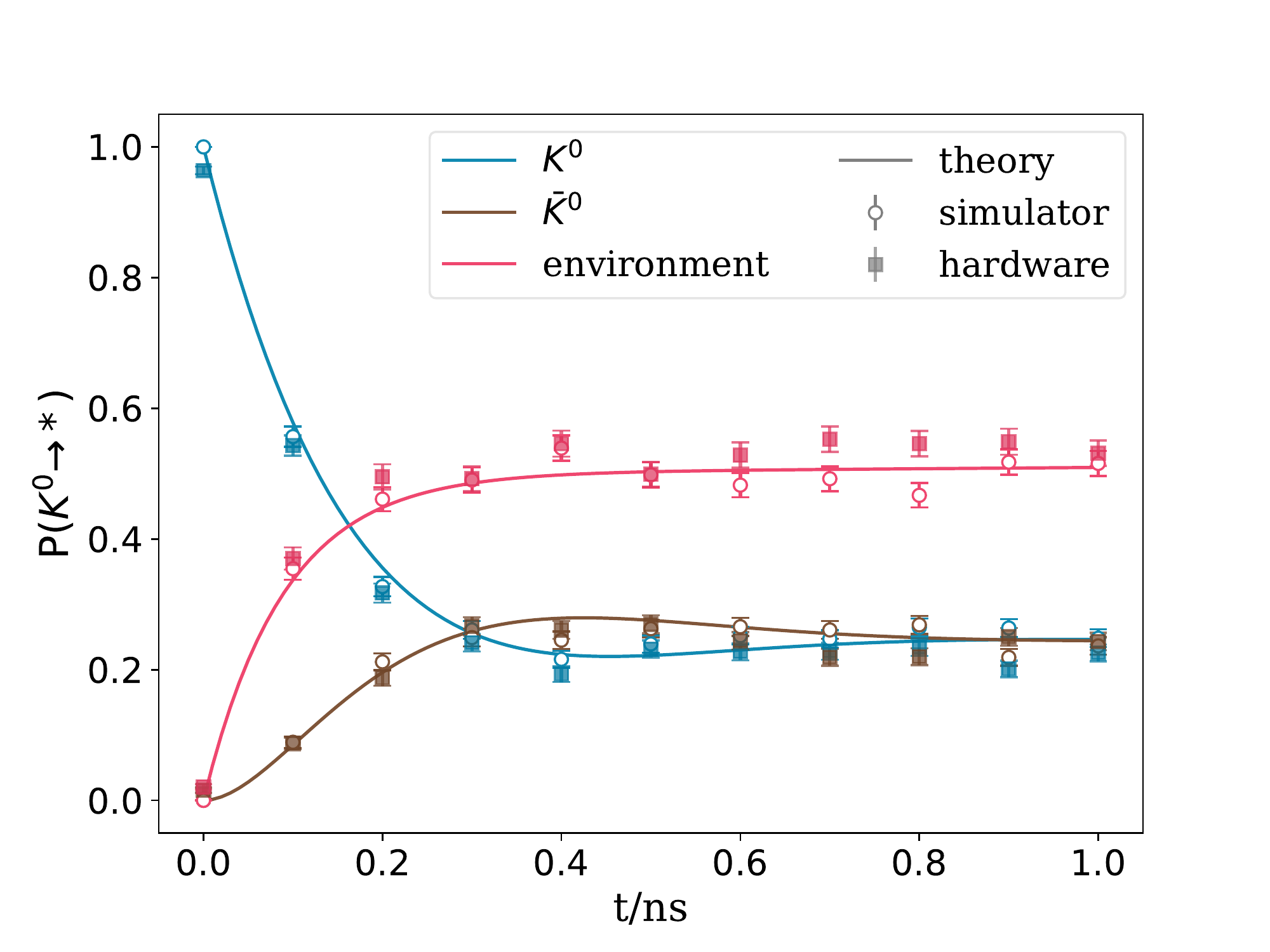}
	\caption{The probabilities $P(K^0 \to *)$ with ``$*$" being $K^0$, $\overline{K}^0$ or the environment in the CP-conserving case, where the solid curves stand for the theoretical calculations, the unfilled circles for the results from a simulator, and the filled squares for those from the hardware.}
	\label{fig:cpc_damping}
	\end{center}
\end{figure}

From the above discussions, it is now evident how the quantum circuit for the evolution operator ${\sf E}(t)$ can be constructed. The circuit diagram has been shown in Fig.~\ref{fig:cpc_circuit}, where the publicly available Quantikz package has been used~\cite{Kay20}. In the assumption of CP conservation, the relevant parameters for the neutral-kaon system include $\Gamma$, $M^{}_{12}$ and $\Gamma^{}_{12}$. Taking the best-fit values of the masses and decay widths from Particle Data Group~\cite{Zyla:2020zbs}, we find $\Gamma = (\Gamma^{}_{\rm S} + \Gamma^{}_{\rm L})/2 = 5.592 \times 10^{9} ~ \rm{s^{-1}}$, $M^{}_{12} = (M^{}_{\rm S} - M^{}_{\rm L})/2 = -2.6465 \times 10^{9} ~ \rm{s^{-1}}$, and $\Gamma^{}_{12} = (\Gamma^{}_{\rm S} - \Gamma^{}_{\rm L})/2 = 5.573 \times 10^{9} ~ \rm{s^{-1}}$. In our calculations, we have used these values and set the initial state of the neutral-kaon system to $|K^0\rangle$. We run the circuit for 1024 shots for each given time $t$ on a quantum simulator and a quantum hardware of IBM. By counting the number of outcomes, we can reconstruct the probabilities of $|K^0\rangle$ transforming to $|K^0\rangle$, $|\overline{K}^0\rangle$ or the environment at the elapsed time $t$. The final numerical results are shown in Fig.~\ref{fig:cpc_damping}, and the theoretical results are also presented for comparison.

Some comments on the results in Fig.~\ref{fig:cpc_damping} are helpful. First of all, the numerical results of the survival $P(K^0 \to K^0)$ and transition $P(K^0 \to \overline{K}^0)$ probabilities from the quantum simulator (unfilled circles) are well compatible with the theoretical expectations (solid curves). Second, as for the results from the IBM hardware, for which the detailed description has been given in the Appendix,  we can observe a little bit-flip effect. The bit-flip effect means that there always exists an error for a physical qubit to be recorded as $|0\rangle$, even when it is actually $|1\rangle$, or vice versa.  This effect may turn an outcome such as $|1^{}_{\rm s}\rangle \otimes |0^{}_{\rm b}\rangle$ into $|1^{}_{\rm s}\rangle \otimes |1^{}_{\rm b}\rangle$, which could change the ratio between these two states. Such an effect can explain the slight discrepancy between the results from the hardware and theoretical expectations.

As we have seen, our strategy for simulating the $K^0$-$\overline{K}^0$ system works very well. All the simulation results are consistent with theoretical calculations within the errors. It is worthwhile to point out that the error bars on the simulation results contain only the statistical uncertainties. The systematical errors caused by the hardware noises have been omitted, whereas the quantum simulator is ideal without any noises.

\subsection{CP Violation}\label{subsec:CPV}

In the presence of CP violation, the matrix elements $M^{}_{12}$ and $\Gamma^{}_{12}$ turn out to be complex, and the relative phase between them will be physical and important. In this case, the time-evolution operator can be written as
\begin{align}
{\sf E}(t) \equiv \exp(-{\rm i} {\sf H}t) = \exp(-{\rm i}Mt) \cdot \exp(-\Gamma t/2) \cdot \exp \left[-{\rm i}(\alpha {\bm \sigma}^{}_x + \beta {\bm \sigma}^{}_y)t \right] \; ,
\label{eq:ECPV}
\end{align}
where $\alpha \equiv {\rm Re} M^{}_{12} - {\rm i}\,{\rm Re}\Gamma^{}_{12}/2$ and $\beta \equiv {\rm Im} M^{}_{12} + {\rm i}\,{\rm Im}\Gamma^{}_{12}/2$ have been defined, and ${\bm \sigma}^{}_{x, y}$ are the first and second Pauli matrices. From Eq.~(\ref{eq:ECPV}), one can observe that ${\sf E}(t) = {\sf E}^{}_1(t) \cdot {\sf E}^{}_2(t) \cdot {\sf E}^{}_3(t)$, where ${\sf E}^{}_1(t)$ and ${\sf E}^{}_2(t)$ remain the same as in Eq.~(\ref{eq:E1E2}) but ${\sf E}^{}_3(t)$ is very different, namely,
\begin{align}
{\sf E}^{}_3(t) \equiv \exp \left[-{\rm i}(\alpha {\bm \sigma}^{}_x + \beta {\bm \sigma}^{}_y) t \right] = \left(\begin{matrix}
       \cos kt & -(\widehat{\beta} + {\rm i} \widehat{\alpha})\sin kt \\
       +(\widehat{\beta} - {\rm i} \widehat{\alpha}) \sin kt & \cos kt
       \end{matrix}\right) \; ,
\label{eq:E3CPV}
\end{align}
with $k \equiv \sqrt{\alpha^2 + \beta^2}$, $\widehat{\alpha} \equiv \alpha/k$ and $\widehat{\beta} \equiv \beta/k$ being complex parameters. Since ${\sf E}^{}_3(t)$ is non-unitary, it cannot be simply represented by any basic logical gates.

In order to overcome this difficulty, we first perform the singular value decomposition (SVD) of ${\sf E}^{}_3(t)$ and obtain ${\sf E}^{}_3= {\sf U} \cdot \widehat{\sf E}^{}_3 \cdot {\sf V}^\dagger$, where ${\sf U}$ and ${\sf V}$ are two $2\times 2$ unitary matrices and $\widehat{\sf E}^{}_3 = {\rm Diag}\{s^{}_1, s^{}_2\}$ is diagonal. After the SVD procedure, we arrive at ${\sf E} = {\sf E}^{}_1 \cdot {\sf U} \cdot {\sf N} \cdot {\sf V}^\dagger$, where ${\sf E}^{}_2$ is proportional to the identity matrix and thus can be combined with $\widehat{\sf E}^{}_3$ into
\begin{align}
{\sf N} \equiv {\sf E}^{}_2 \cdot \widehat{\sf E}^{}_3 =        \left(\begin{matrix}s^{}_{1}  e^{-\Gamma t/2} & 0 \\ 0 & s^{}_{2}e^{-\Gamma t/2} \end{matrix}\right) \; .
\label{eq:N}
\end{align}
It is evident that we need to encode the matrices ${\sf U}$, ${\sf N}$ and ${\sf V}^\dagger$ with quantum gates. Notice that the argument $t$ has been suppressed for all these matrices. For two unitary matrices ${\sf U}$ and ${\sf V}^\dagger$, we implement the so-called $\bf U$ gate from the IBM Qiskit (used to be ${\bf U3}$ gate before Qiskit 0.16.0)
\begin{align}
{\bf U}(\theta,\phi,\lambda) = \left(\begin{matrix}
\cos(\theta/2) & -e^{i\lambda}\sin(\theta/2) \\
e^{i\phi}\sin(\theta/2) & e^{i\lambda+i\phi}\cos(\theta/2)
\end{matrix}\right) \; ,
\label{eq:U}
\end{align}
where the corresponding parameters $(\theta, \phi, \lambda)$ can be determined from ${\sf U}$ and ${\sf V}^\dagger$ up to some irrelevant overall phases. The details of the SVD procedure and the explicit expressions of ${\sf U}$, $\widehat{\sf E}^{}_3$ and ${\sf V}^\dagger$ are given in the Appendix. Note that one can in principle replace the basic gates, like $\bf H$, $\bf X$ and $\bf Rz$ in our circuits, by the more general $\bf U$ gate. However, for convenience, we shall use those traditional gates in the first place, and implement the $\bf U$ gate only when there are no alternatives.

\begin{figure}[!t]
	\begin{center}
    	\begin{tikzpicture}
        	\node[scale=0.95] {
            	\begin{quantikz}
                	\lstick{$q^{}_{\rm s}$} &\ket{0} &\gateU{\theta,\phi,\lambda} &\ctrl{1}\gategroup[wires=2,steps=4,style={dashed,rounded corners,fill=blue!20,inner xsep=2pt},background]{${\sf G}$}
                	               &\gateX &\ctrl{1}             &\gateX &\gateU{\theta',\phi',\lambda'} &\meas{} \\
                	\lstick{$q^{}_{\rm b}$} &        &\gateO    &\gateRy{2\arccos b}
                	               &\qw    &\gateRy{2\arccos a}  &\qw    &\qw       &\meas{}
    	        \end{quantikz}
        	};
	    \end{tikzpicture}
\caption{Quantum circuit for the time evolution of the $K^0$-$\overline{K}^0$ system in the CP-violating case, where the circuit is initialized in $|K^0\rangle$ state. Note that the leftmost $\bf U$ gate is for ${\sf V}^\dagger$ while the rightmost one is for ${\sf U}$.}
	    \label{fig:cpv_circuit}
	\end{center}
\end{figure}
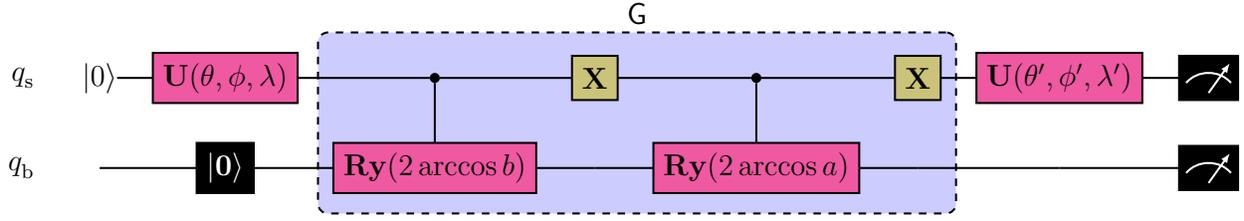

The last step is to encode the diagonal and non-unitary matrix ${\sf N}$ in Eq.~(\ref{eq:N}). As we have learned from the CP-conserving case, such a non-unitary matrix can be realized by adding one qubit for the environment and constructing a $4\times 4$ unitary matrix ${\sf G}$. One can prove that the two non-zero elements of ${\sf N}$, which can be identified as $a = s^{}_1\exp{(-\Gamma t/2)}$ and $b = s^{}_2\exp{(-\Gamma t/2)}$, are both positive and less than one for $t>0$. Consequently, the unitary matrix ${\sf G}$ in Eq.~(\ref{eq:G}) and its realization in Eq.~(\ref{eq:Gcode}) are also applicable to the present case but now with different functions $a(t)$ and $b(t)$. The quantum circuit in the CP-violating case has been depicted in Fig.~\ref{fig:cpv_circuit}, and will be used to simulate the time evolution of the $K^0$-$\overline{K}^0$ system both on a quantum simulator and on the IBM hardware.

\begin{figure}[!t]
	\begin{center}
	\includegraphics[scale=0.65]{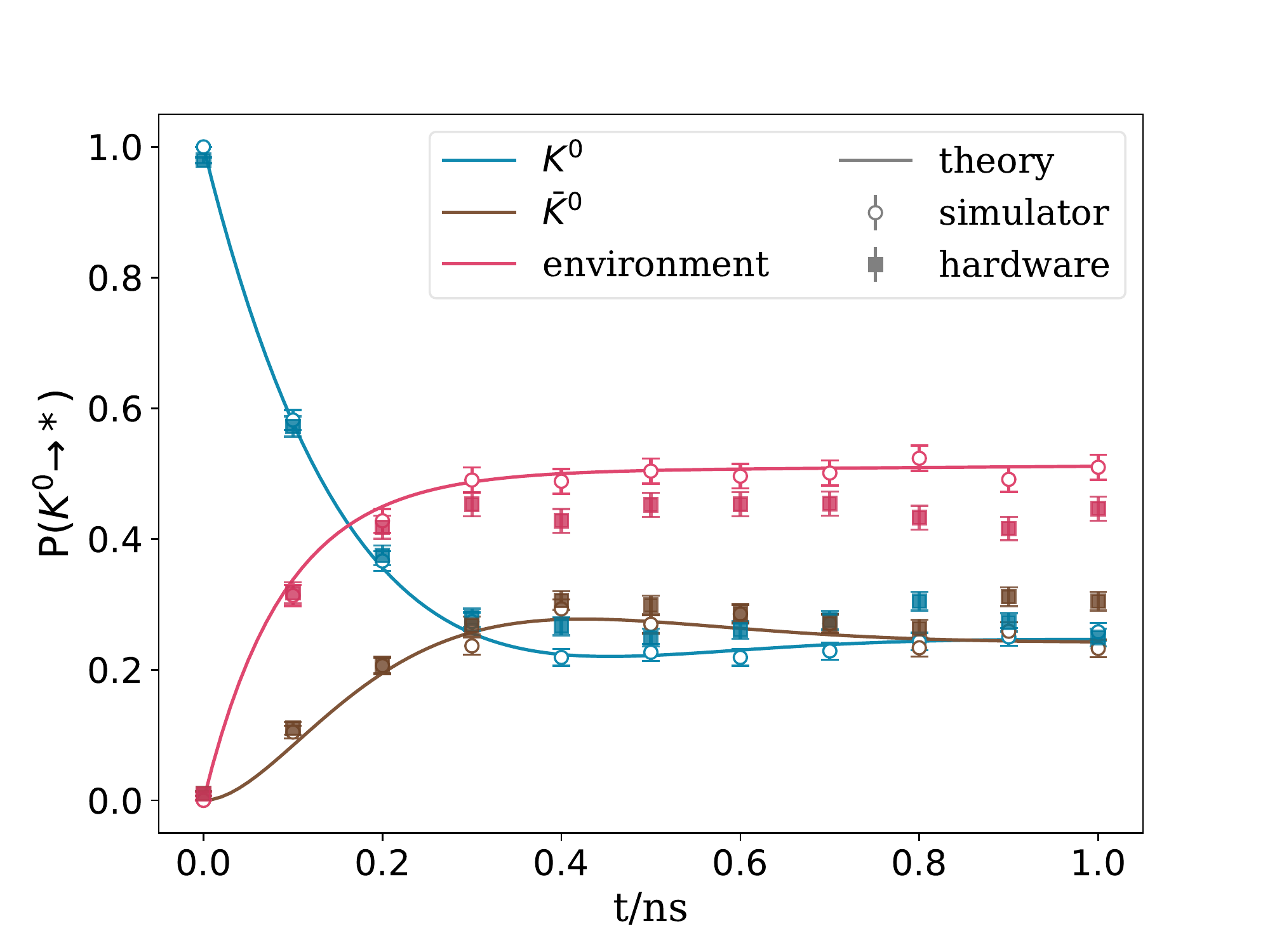}
	\caption{The probabilities $P(K^0 \to *)$ with ``$*$" being $K^0$, $\overline{K}^0$ or the environment in the CP-violating case, where the solid curves stand for the theoretical calculations, the unfilled circles for the results from a simulator, and the filled squares for those from the hardware. The realistic phase difference $\phi^{}_{\rm M} - \phi^{}_{\Gamma} = 180.4^\circ$ has been input.}
	\label{fig:cpv_damping}
	\end{center}
\end{figure}

\begin{figure}[!t]
	\begin{center}
	\includegraphics[scale=0.65]{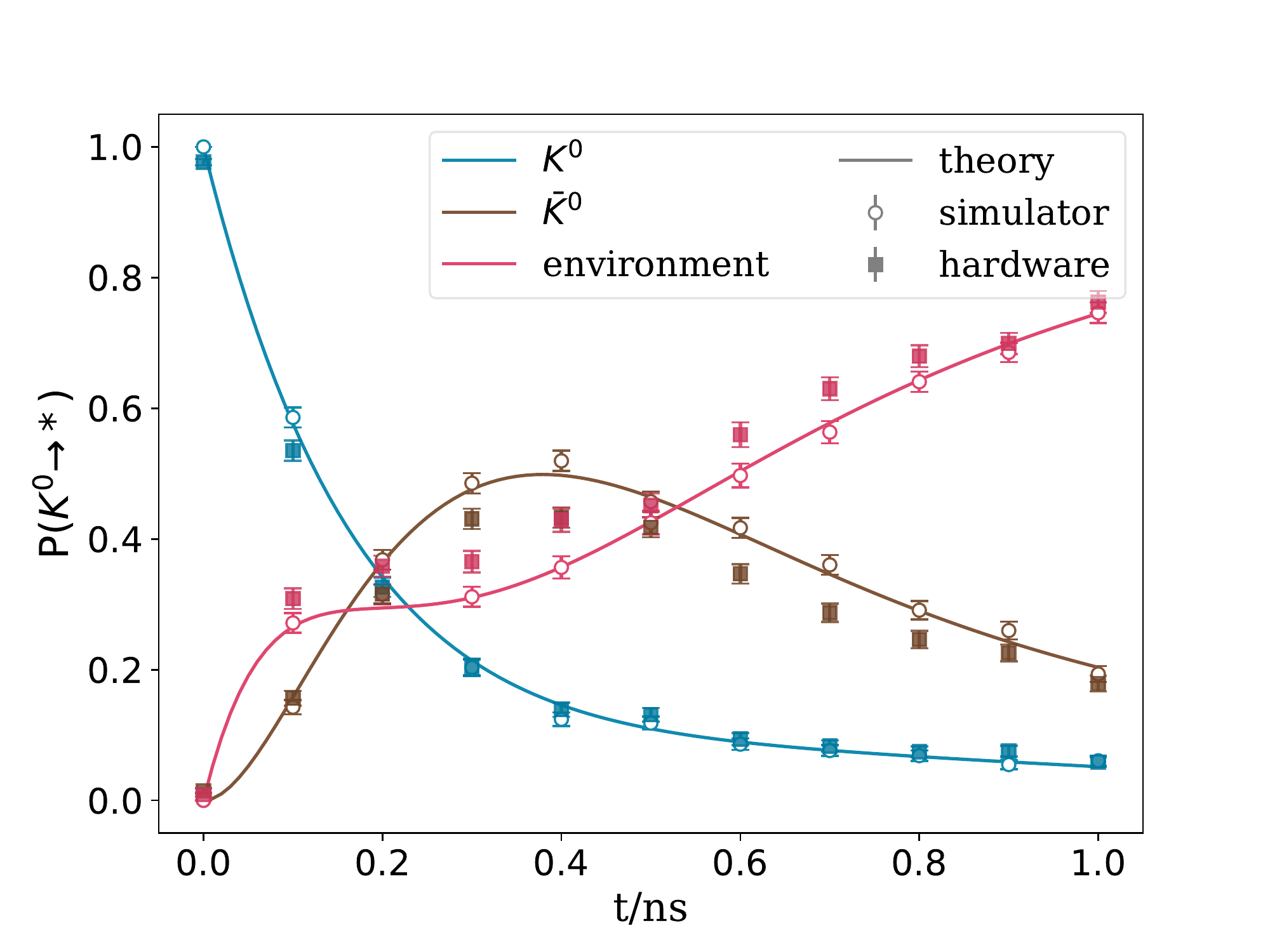}
	\caption{The probabilities $P(K^0 \to *)$ with ``$*$" being $K^0$, $\overline{K}^0$ or the environment in the CP-violating case, where the solid curves stand for the theoretical calculations, the unfilled circles for the results from a simulator, and the filled squares for those from the hardware. The unrealistic phase difference $\phi^{}_{\rm M} - \phi^{}_{\Gamma} = 60^\circ$ has been input, while the other parameters are fixed as in Fig.~\ref{fig:cpv_damping}.}
	\label{fig:cpv_damping60}
	\end{center}
\end{figure}

Unlike the CP-conserving case with three relevant parameters, we now need four. In addition to the average decay width $\Gamma = 5.592 \times 10^{9} ~ \rm{s^{-1}}$, we have $\left|\Gamma^{}_{12}\right| = 5.573 \times 10^{9} ~ \rm{s^{-1}}$, $\left|M^{}_{12}\right| = 2.647 \times 10^{9} ~ \rm{s^{-1}}$ and $\phi^{}_{\rm M} - \phi^{}_\Gamma = 180.4^\circ$. Note that one is allowed to freely choose the value of $\phi^{}_{\rm M}$ or $\phi^{}_\Gamma$ while keeping their difference unchanged. With these input parameters, we have run the simulations and compared the numerical results with theoretical calculations. The final results have been presented in Fig.~\ref{fig:cpv_damping}. Some important observations can be made.

First, as in the previous CP-conserving case, the numerical results from the quantum simulator (unfilled circles) are perfectly consistent with the theoretical expectations (solid curves). However, the results from the quantum hardware (filled squares) show some bias. This time, the most likely cause is the bit-flip effect that changes $|1^{}_{\rm s}\rangle \otimes |1^{}_{\rm b}\rangle$ or $|0^{}_{\rm s}\rangle \otimes |1^{}_{\rm b}\rangle$ into $|0^{}_{\rm s} \rangle \otimes |0^{}_{\rm b}\rangle$ or $|1^{}_{\rm s} \otimes |0^{}_{\rm b}\rangle$, respectively. Since we have identified $|1^{}_{\rm s}\rangle \otimes |1^{}_{\rm b}\rangle$ and $|0^{}_{\rm s} \otimes |1^{}_{\rm b}\rangle$ in the final measurements as the environment states and the remaining two as $|K^0\rangle$ and $|\overline{K}^0\rangle$, an outcome of environment states may transform into the $K^0$-$\overline{K}^0$ system if some digit in the environment state drops from $|1\rangle$ to $|0\rangle$. This results in an increase of the probabilities for $|K^0\rangle$ or $|\overline{K}^0\rangle$, as one can see in Fig.~\ref{fig:cpv_damping}. As the bit-flip effect is intimately related with the hardware properties, the observed bias can be attributed to the hardware noises, which are not beyond one's expectation.

Second, in comparison with the results in Fig.~\ref{fig:cpc_damping} in the CP-conserving case, one can conclude that there is no much difference between CP-conserving and CP-violating cases. This is because the strength of CP violation is quite small, namely, $\sin(\phi^{}_{\rm M} - \phi^{}_\Gamma) \approx 0$, which can be safely neglected. In order to examine the impact of CP violation on the probabilities and the performance of quantum computations, we have also redone the calculations by inputting an artificially large CP-violating phase, i.e., $\phi^{}_{\rm M} - \phi^{}_\Gamma = 60^\circ$, while keeping the other input parameters unchanged. The corresponding results are shown in Fig.~\ref{fig:cpv_damping60}. Compared to the case of small CP violation, the remarkable difference is the enhanced leakage probability to the environment. As before, the outcome from the ideal quantum simulator matches the theoretical expectation very well. Still, there is a noticeable bias between the theory and the quantum hardware. The probability for the final-state $|\overline{K}^0\rangle$ is much lower than the expected value, whereas that for the environment state just appears to be the opposite, namely, with a higher probability than the theoretical prediction. The most likely reason is the bit-flip effect that transforms the outcome of $|1^{}_{\rm s} \otimes |0^{}_{\rm b}\rangle$ into $|1^{}_{\rm s}\rangle \otimes |1^{}_{\rm b}\rangle$, increasing the probability of the environment state and reducing the probability of $|\overline{K}^0\rangle$.

\section{Summary}\label{sec:sum}

Motivated by recent tremendous progress in the developments of quantum algorithms and the actual implementation of quantum computing devices, we have investigated possible applications to elementary particle physics. Different from the neutrino flavor oscillations considered in Ref.~\cite{Arguelles:2019phs}, the neutral-kaon system is essentially an open system and the non-Hermitian effective Hamiltonian for the two-dimensional Hilbert space of $|K^0\rangle$ and $|\overline{K}^0\rangle$ states implies a non-unitary time evolution. Concentrating on the neutral-kaon oscillations $K^0 \leftrightarrow \overline{K}^0$, we have proposed a feasible design of the quantum circuits in both CP-conserving and CP-violating cases.

The strategy to design the quantum circuits is essentially to embed the non-unitary evolution of the two-level system into the unitary evolution of a larger system comprised of the principal system under study and the environment. This goal can be achieved by introducing another qubit for the environment, and the whole system can be projected into the subsystem of interest by choosing a particular quantum state of the environment. Although the neutral-kaon system with CP violation is slightly complicated by the non-unitary time-evolution matrix, to which the SVD procedure should be applied in the first place, our strategy is universally working for CP-conserving and CP-violating cases. The numerical calculations on the quantum simulator and the quantum hardware have been carried out to demonstrate the validness and efficiency of the proposed quantum circuits.

It is worth stressing that although only the neutral-kaon system is examined, the method should be equally applicable to other neutral-meson systems, such as $D^0$-$\overline{D}^0$ and $B^0$-$\overline{B}^0$ oscillations. In the latter cases, the masses and decay widths of two energy eigenstates will be quite different. The treatment of non-unitary time evolution can also be generalized to deal with other open systems. In the near future, we hope to extend the neutral-kaon system by incorporating one or more quantum states, into which the neutral kaons can decay. Such a scenario is interesting since it is much closer to the reality in nature. Certainly, the quantum algorithm for this extended system must be more involved and deserves further investigations.

\section*{Acknowledgements}

The authors are indebted to Prof. Zhi-zhong Xing for helpful discussions and valuable suggestions. This work was supported in part by the National Natural Science Foundation of China under Grant No.~11775232, No.~11835013, No.~11935017 and No.~12070131001 (CRC 110 by DFG and NNSFC), and by the CAS Center for Excellence in Particle Physics. The use of IBM Quantum services in this work is greatly acknowledged. The views expressed are those of the authors, and do not reflect the official policy or position of IBM or the IBM Quantum team.

\section*{Appendix}

\subsection*{a. IBM Quantum Computers}

The quantum hardware that we used is the five-qubit machine {\it ibm\_santiago}.  The simulations were carried out on two physical qubits labelled as 0 and 1, which are connected directly in the processor's topology diagram. The average T1 and T2 time for the processor were reported as 131.13~$\mu$s and 122.79~$\mu$s, when this paper was in preparation, and the average readout error was 2.04\% while the average CNOT error was 1.02\%.  The machine is regularly calibrated to keep it in a good condition, so the parameters mentioned above may vary with time when the calibration is performed. This is manifested to some extent in the different bit-flip effects in the simulation results for different CP-violating phases in Fig.~\ref{fig:cpv_damping} and Fig.~\ref{fig:cpv_damping60}, which were obtained by different calculations on the machine.

Our quantum program was coded in Python with the package Qiskit 0.23.0, and the quantum simulator is a part of Qiskit. The computations on the simulator were executed on a local device. We take the results from the simulator as a benchmark and compare them with those from the real quantum computer. The differences between the results in these two cases can be attributed to the noises of the quantum devices.

\subsection*{b. Singular Value Decomposition}

Now we present some calculational details for the singular value decomposition (SVD) of the evolution matrix ${\sf E}^{}_3(t)$ in \eqref{eq:E3CPV}, which is rewritten as
\begin{equation}
{\sf E}_3(t)=\left(\begin{matrix}\cos kt & -(\widehat{\beta}+{\rm i}\widehat{\alpha})\sin kt \cr +(\widehat{\beta}-{\rm i}\widehat{\alpha})\sin kt & \cos kt\end{matrix}\right)\equiv \left(\begin{matrix} c & -b-{\rm i}a \cr
b-{\rm i}a & c \end{matrix}\right) \; ,
\label{eq:E3abc}
\end{equation}
where three complex parameters $a \equiv \widehat{\alpha}\sin kt$, $b \equiv \widehat{\beta} \sin kt$, and $c \equiv \cos kt$ have been defined for later convenience. As has been mentioned in the main text, the SVD of the matrix ${\sf E}^{}_3(t)$ is given by ${\sf E}^{}_3(t) = {\sf U} \cdot \widehat{\sf E}^{}_3 \cdot {\sf V}^\dagger$, where ${\sf U}$ and ${\sf V}$ are unitary matrices and $\widehat{\sf E}^{}_3 = {\rm Diag}\{s^{}_1, s^{}_2\}$ is a diagonal matrix with two singular values $s^{}_{1,2}$ being real and non-negative. The strategy to find the unitary matrices ${\sf U}$ and ${\sf V}$, as well as the singular values $s^{}_{1, 2}$, is quite standard. First of all, one can look for two unitary matrices ${\sf u}$ and ${\sf v}$ such that ${\sf E}^{}_3 = {\sf u} \cdot {\sf D} \cdot {\sf v}^\dagger$ with ${\sf D} \equiv {\rm Diag}\{\lambda^{}_1, \lambda^{}_2\}$, where $\lambda^{}_1$ and $\lambda^{}_2$ are complex. For this purpose, we construct two Hermitian matrices
\begin{equation}
{\sf G}^{}_u \equiv {\sf E}^{}_3 \cdot {\sf E}^\dagger_3 \; , \quad {\sf G}^{}_v \equiv {\sf E}^\dagger_3 \cdot {\sf E}^{}_3 \; ,
\label{eq:GuGv}
\end{equation}
which can be diagonalized via ${\sf G}^{}_u = {\sf u} \cdot |{\sf D}|^2 \cdot {\sf u}^\dagger$ and ${\sf G}^{}_v = {\sf v} \cdot |{\sf D}|^2 \cdot {\sf v}^\dagger$, respectively. The diagonal matrix $|{\sf D}|^2 \equiv {\rm Diag}\{|\lambda^{}_1|^2, |\lambda^{}_2|^2\}$ can be identified with $\widehat{\sf E}^2_3$, for which the relations $s^{}_1 = |\lambda^{}_1|$ and $s^{}_2 = |\lambda^{}_2|$ hold. On the other hand, one can immediately verify that $\{{\sf u}, {\sf v}\}$ are related to $\{{\sf U}, {\sf V}\}$ by a diagonal phase matrix, namely,
\begin{eqnarray}
{\sf U} = {\sf u} \cdot \left( \begin{matrix} e^{+{\rm i}\theta^{}_1/2} & 0 \cr 0 & e^{+{\rm i}\theta^{}_2/2}\end{matrix} \right) \;, \quad {\sf V} = {\sf v} \cdot \left( \begin{matrix} e^{-{\rm i}\theta^{}_1/2} & 0 \cr 0 & e^{-{\rm i}\theta^{}_2/2}\end{matrix} \right) \; ,
\label{eq:UuVv}
\end{eqnarray}
where $\lambda^{}_1 = s^{}_1 e^{{\rm i}\theta^{}_1}$ and $\lambda^{}_2 = s^{}_2 e^{{\rm i}\theta^{}_2}$. Then, the remaining task is to find the unitary matrices ${\sf u}$ and ${\sf v}$, as well as the singular values $s^{}_{1, 2}$, by diagonalizing the Hermitian matrices ${\sf G}^{}_u$ and ${\sf G}^{}_v$, and calculate ${\sf u}^\dagger \cdot {\sf E}^{}_3 \cdot {\sf v} = {\sf D}$ in order to extract the phases $\theta^{}_{1,2}$ from $\lambda^{}_{1,2}$. More details of this procedure can be found below.
\begin{itemize}
\item With the help of Eqs.~(\ref{eq:E3abc}) and (\ref{eq:GuGv}), we can obtain the independent matrix elements of ${\sf G}^{}_u$ and ${\sf G}^{}_v$, i.e.,
\begin{eqnarray}
{\sf G}_u^{11} &=& {\sf G}_v^{22} = |a|^2 + |b|^2 + |c|^2 - 2\, {\rm Im}(ab^*) \; , \nonumber \\
{\sf G}_u^{22} &=& {\sf G}_v^{11} = |a|^2 + |b|^2 + |c|^2 + 2\, {\rm Im}(ab^*) \; , \nonumber \\
{\sf G}_u^{12} &=& {\sf G}_v^{12} = 2\, {\rm Im}(ac^*) - 2{\rm i}\, {\rm Im}(bc^*) \; .
\label{eq:Guv}
\end{eqnarray}
It is easy to observe from Eq.~(\ref{eq:Guv}) that these two Hermitian matrices ${\sf G}^{}_u$ and ${\sf G}^{}_v$ are related by the $\mathcal{P}\mathcal{T}$ transformation ${\sf G}^{}_u = (\mathcal{P}\mathcal{T}) \cdot{\sf G}^{}_v \cdot (\mathcal{P}\mathcal{T})^{-1}$, where
\begin{equation}
\mathcal{P}\mathcal{T} \equiv \left(\begin{matrix} 0 & 1 \cr 1 & 0 \end{matrix} \right) \cdot \mathcal{K} \; ,
\label{eq:PT}
\end{equation}
with $\mathcal{K}$ being the ordinary complex-conjugate operator~\cite{Bender:2007nj, Ohlsson:2019noy}. Once ${\sf G}^{}_v$ is diagonalized, ${\sf G}^{}_u$ will be diagonalized accordingly. By solving the secular equation of the Hermitian matrix ${\sf G}^{}_v$, namely, ${\sf G}^{}_v v^{}_i = s^2_i v^{}_i$ (for $i = 1, 2$),
one obtains the eigenvalues
\begin{equation}
s^2_{1,2} = \frac{1}{2}\left[({\sf G}_v^{11}+{\sf G}_v^{22})\pm \sqrt{({\sf G}_v^{11}-{\sf G}_v^{22})^2+4|{\sf G}_v^{12}|^2}\right] \; .
\label{eq:eigenvalues}
\end{equation}
Meanwhile, the corresponding eigenvectors $v^{}_1 \equiv (v^1_1, v^2_1)^{\rm T}$ and $v^{}_2 \equiv (v^1_2, v^2_2)^{\rm T}$ can be explicitly figured out
\begin{eqnarray}
\left(\begin{matrix}v_1^1\cr v_1^2\end{matrix}\right) &=& \left[\frac{|{\sf G}_v^{12}|^2}{|{\sf G}_v^{12}|^2 + (s^2_1 - {\sf G}_v^{11})^2}\right]^{1/2} \left(\begin{matrix}1\cr (s^2_1 - {\sf G}_v^{11})/{\sf G}^{12}_v \end{matrix}\right) \; , \nonumber \\
\left(\begin{matrix}v_2^1 \cr v_2^2\end{matrix}\right) &=& \left[\frac{|{\sf G}_v^{21}|^2}{|{\sf G}_v^{21}|^2 + (s^2_2 - {\sf G}_v^{22})^2}\right]^{1/2}\left(\begin{matrix} (s^2_2 - {\sf G}_v^{22})/{\sf G}^{21}_v \cr 1\end{matrix}\right) \; ,
\label{eq:vij}
\end{eqnarray}
where $v^1_1$ and $v^2_2$ are chosen to be real and the normalization conditions $(v^1_1)^2 + |v^2_1|^2 = 1$ and $|v^1_2|^2 + (v^2_2)^2 = 1$ have been used. Consequently, the unitary matrix ${\sf v} = (v^{}_1, v^{}_2)$ is obtained.

\item Inserting ${\sf G}^{}_v = (\mathcal{P}\mathcal{T})^{-1} \cdot {\sf G}^{}_u \cdot (\mathcal{P}\mathcal{T})$ into the eigen equations ${\sf G}^{}_v v^{}_i = s^2_i v^{}_i$, one arrives at
\begin{equation}
 {\sf G}^{}_u (\mathcal{P}\mathcal{T}) v^{}_i = s^2_i (\mathcal{P}\mathcal{T}) v^{}_i \; ,
 \label{eq:PTv}
\end{equation}
from which the eigenvectors $u^{}_i$ of ${\sf G}^{}_u$ can be readily read out as
\begin{equation}
u^{}_i = (\mathcal{P}\mathcal{T}) v^{}_i \; ,
\label{eq:ui}
\end{equation}
and thus the unitary matrix ${\sf u} = (u^{}_1, u^{}_2)$ that diagonalizes ${\sf G}^{}_u$. It is straightforward to verify that the previously obtained unitary matrices ${\sf u}$ and ${\sf v}$ really diagonalize ${\sf E}^{}_3$ via ${\sf u}^\dagger \cdot {\sf E}^{}_3 \cdot v = {\sf D}$. Finally, we get ${\sf U}$ and ${\sf V}$ from Eq.~(\ref{eq:UuVv}).
\end{itemize}

\end{document}